\documentclass[article,aps,pra,twocolumn,showpacs,superscriptaddress, nofootinbib ]{revtex4}  
\usepackage{graphicx}  
\usepackage{dcolumn}   
\usepackage{bm}        
\usepackage{amssymb}   
\usepackage{amsmath}
\usepackage{braket}
\usepackage[caption=false]{subfig}
\usepackage{verbatim}
\usepackage{tikz}
\usetikzlibrary{quantikz}
\usepackage{eqnarray,amsmath}

\usepackage{color}
\newcommand{\be}{\begin{equation}}
\newcommand{\ee}{\end{equation}}
\newcommand{\ben}{\begin{eqnarray}}
\newcommand{\een}{\end{eqnarray}}
\newcommand{\bes}{\begin{subequations}}
\newcommand{\ees}{\end{subequations}}
\newcommand{\bF}{\begin{figure}}
\newcommand{\eF}{\end{figure}}

\newcommand{\RNum}[1]{\uppercase\expandafter{\romannumeral #1\relax}}

\newcommand{\ketbra}[2]{\left|#1\right\rangle\!\left\langle#2\right|}

\begin{document}
\title{Exponential speedup in measuring {out-of-time-ordered} correlators with a single bit of quantum information}

	\author{Sreeram PG}
	\affiliation{Department of Physics, Indian Institute of Technology Madras, Chennai, India 600036}
	
	\author{Naga Dileep Varikuti}
	\affiliation{Department of Physics, Indian Institute of Technology Madras, Chennai, India 600036}
	
	\author{Vaibhav Madhok}
	     \email{vmadhok@gmail.com}
	\affiliation{Department of Physics, Indian Institute of Technology Madras, Chennai, India 600036}

\begin{abstract}
{Out-of-time-ordered} correlators (OTOC) are a quantifier of quantum information scrambling and  quantum chaos. We propose an efficient quantum algorithm to measure OTOCs that provides an exponential {speed-up} over the best known classical algorithm provided the OTOC operator to be estimated admits an efficient gate decomposition. We also discuss a  scheme to obtain information about the eigenvalue spectrum and the spectral density of OTOCs.
\end{abstract}
\maketitle

\section{Introduction}
Connections between non-integrability, many-body physics, complexity, ergodicity, and entropy generation are the cornerstones of statistical mechanics. The aim of quantum chaos is to extend these questions in the quantum domain.
Foundational works in this regard include semiclassical methods connecting classical periodic orbits to {the} density of states
 level statistics \cite{Berry375}, properties of Wigner functions \cite{Berry77a}, and quantum scars in ergodic phase spaces \cite{heller1984bound} and connections to random matrix theory.
Search for these footprints of chaos, and characterization of ``true" quantum chaos, independent of any classical limit, has important consequences both from a foundational point of view as well for quantum information processing. 
For example, such studies address complexity in quantum systems and play a potentially crucial role in information processing protocols like quantum simulations that are superior to their classical counterparts. 

Characterization of chaos in the quantum domain has been much contested since, unlike its classical counterpart, unitary quantum evolution preserves the overlap between two initial state vectors and hence rules out hypersensitivity to initial conditions. However, a deeper study reveals chaos in quantum systems. 
These issues have been extensively studied in the last few decades and several quantum signatures of classical chaos have been discovered. This interestingly coincides with exquisite control of individual quantum systems in the laboratory and the ability to coherently drive these systems with non-integrable/chaotic Hamiltonians. Recent trends include studies involving connections of quantum chaos to {out-of-time-ordered} correlators (OTOC) and the rate of scrambling of quantum information in many-body systems with consequences ranging from the foundations of quantum statistical mechanics, quantum phase transitions, and thermalization on the one hand to information scrambling inside a black hole on the other hand \cite{manybody1, manybody2, manybody3, manybody4, qgravity1, chaos1,chaos2, shock1, pawan, shenker2, shenker3, kitaev1, kitaev2}.

OTOCs have been much talked about in the quantum information circle recently and a number of ways to measure OTOCs have been proposed including a protocol employing an interferometric scheme in cold atoms \cite{swingle}.  An alternative method involving two-point projective measurements was proposed \cite{campisi}, giving a scheme for the measurement of OTOCs using the two-point measurement scheme, developed in the field of non-equilibrium quantum thermodynamics  elucidating the connections between information scrambling and thermodynamics.
Various other protocols are reported in \cite{meas1,meas2,meas3}. Measuring OTOCs in experiments is not easy, as the implementation of perfect time reversal in an experimental setting is impossible because of dissipation. However experimental implementation has been achieved in some systems. Measurement of OTOCs for an Ising spin chain in an NMR simulator has been reported \cite{nmr,wei}. A many-body time-reversal protocol using trapped ions has been proposed and demonstrated \cite{garttner} which though universal is not scalable.
These experiments measure infinite temperature OTOCs, an observation that will be important for us.

 { In order to explore any quantum signatures of chaos, one has to numerically process data structures whose computational complexity scale 
	exponentially with the number of qubits required to simulate the system. 
	In this paper, we give a quantum algorithm that gives an exponential {speed-up} in measuring OTOCs provided that the number of gates, $K$, required in the decomposition of the times evolution operator of the system scales \textit{polynomially} with $n$, where $n$ is the number of qubits used in the implementation and, $N$, the dimension of the Hilbert space with $N = 2^n$. This implies that the algorithm measures the OTOCs in a time that scales as poly(n), which is exponentially faster than any classical algorithm. Our algorithm is based on the Deterministic Quantum Computation with one pure qubit (DQC1) algorithm, which is the first mixed state scheme of quantum computation. Therefore, this can be naturally implemented by a high-temperature NMR based quantum information processor.  It involves a deterministic quantum control of one qubit model, using scattering circuit\cite{knill,scattering}. This algorithm is also called the `power of one qubit' as the main primary resource required for this algorithm is one pure qubit. Moreover, the essential part of simulations, state initialization, and readout, that are often quite involved in certain models of quantum computation \cite{van2001powerful}.  We give a quantum circuit to evaluate OTOCs—which bypasses the need to prepare a complex initial state and can be accomplished by a very simple measurement.
	Applications include estimation of fidelity decay and density of states in quantum chaos \cite{exponential, PhysRevA.68.022302}, computing Jones {polynomials} from knot theory \cite{shor2007estimating, jones2004nuclear} and phase estimation in quantum metrology \cite{PhysRevA.77.052320}. Although the DQC1 model of quantum information processing (QIP) is believed to be less powerful than a universal quantum computer, its natural implementation in high-temperature NMR makes it an ideal candidate for probing OTOCs and mixed state quantum computation protocols.}

\section{Out-of-time-ordered correlators (OTOCs)}

OTOC were first proposed by Larkin and Ovchinnikov in the context of semiclassical approximations in the theory of superconductivity \cite{larkin}. They later reemerged in the study of many-body systems \cite{manybody1,manybody2, manybody3, manybody4} quantum gravity \cite{qgravity1} and  quantum chaos \cite{chaos1,chaos2, shock1, pawan, shenker2, shenker3, kitaev1, kitaev2}. In quantum information literature, OTOC is used as a  probe to study the dynamics of information. One can probe the macroscopic irreversibility of the dynamics,  the spread of quantum information from a localized point to the rest of the system via entanglement and correlations, and also the aspects of thermalization \cite{unscrambling,scrambling1, scrambling2}. Consider a chain of interacting spins. Then a correlator of two operators acting at two different sites can be defined as
\begin{equation}\label{eqq}
C_{W,V}(\tau)=\dfrac{1}{2}\langle[W(x, \tau), V(y, 0)]^{\dagger}[W(x, \tau), V(y, 0)]\rangle, 
\end{equation}
where the local operators $W$ and $V$ are unitary and/or Hermitian that act on sites $x$ and $y$ respectively and $W(x, \tau)=U^{\dagger}(\tau)W(x, 0)U(\tau)$ is the Heisenberg evolution of operator $W$ under time evolving operator $U(\tau)$. The average is taken with respect to the thermal state at some temperature which we take to be infinite. In particular, if the operators $W$ and $V$ are unitary, the above equation becomes,
\begin{equation}
 C_{W, V}(\tau)=1-\mathtt{Re}\langle W(x,\tau)^{\dagger}V(y, 0)^{\dagger}W(x, \tau)V(y, 0)\rangle.
\end{equation}

In classical physics, the chaos is defined as the sensitive dependence on initial conditions. If we replace $W$ and $V$ in the Eq (\ref{eqq}) with position($Q$) and momentum ($P$) operators, and taking a semi-classical limit, we notice that $\hbar^2\{Q(\tau), P(0) \}^2=\left(\hbar\frac{\delta Q(\tau)}{\delta Q(0)}\right)^{2} \approx \mathrm{exp}(2\lambda \tau)$. The quantum-classical correspondence principle implies that the quantity $C_{W, V}(\tau)$ grows exponentially till the Ehrenfest time($\tau_{Eh}$). However, unlike the classical systems, the lyapunov exponent($\lambda $) calculated from OTOC is bounded by $\frac{2\pi}{\beta}$ \cite{chaos1}. Beyond the $\tau_{Eh}$, the quantum corrections start dominating and the quantum-classical correspondence breaks down. 

An interesting feature of OTOC is that it measures the spreading of initially localized operators across system degrees of freedom as the operator evolves in Heisenberg fashion \cite{pawan,ope1, ope2, ope3, ope4, ope5}. Consider a pair of local operators $W$ and $V$ that act on different subspaces of total Hilbert space($\mathcal{H}$) under a chaotic time evolution $U(\tau)=\exp(iH\tau)$. We assume that the Hamiltonian is generic with local interactions. Under this evolution, the operator $W$ will evolve in time and it can be expanded in Taylor series around $\tau=0$ as 
\begin{eqnarray}\label{eqq2}
 W(\tau)&=&\sum_{n}\dfrac{\tau^n}{n!}\dfrac{d^n W}{d\tau^n}\nonumber\\
&=& W(0)+i\tau[H, W]+(i\tau)^2[H,[H, W]]+...
\end{eqnarray}
This implies that the operators $W(\tau)$ and $V$ in general do not commute for time $\tau \neq 0$. For example, consider one dimensional Ising spin chain with nearest-neighbor interactions. Let $W(i, \tau=0)=\sigma_{z}^i$ acts on site $i$ at time $\tau =0$. On substituting $W$ in second line of the series in the Eq (\ref{eqq2}), the first order commutator will give us the sum of products of local operators acting on the sites $i-1, i$ and $i+1$ i.e $[H, \sigma_{z}^{i}]=f(i-1, i, i+1)$. As time flows, the higher ordered nested commutators also will contribute to the expansion of $W(\tau)$ thus making the quantity $[W(\tau), V]\neq 0$ \cite{shock1}.  

Lieb and Robbinson \cite{lieb1972finite} showed that for short range interacting Hamiltonians, the quantity $C_{W, V}$ is bounded i.e $C_{W,V}(\tau)\leq ce^{-a(i-v\tau)}$. Where $a$ and $c$ are constants and $v$ is called Lieb-Robbinson velocity. This bound on OTOC imply a light-cone like structure in quantum lattice models. 
 it is worthwhile to note that the growth of the OTOC
is a  quantum measure, can be used in systems with no 
obvious classical limits.


\section{Determinstic Quantum Computation with one pure qubit (DQC1)}

{ Single qubit quantum computation, although limited in applicability is interesting from a fundamental point of view. Despite involving minimal entanglement, DQC1 gives an advantage over classical computing. It has been shown that none of the classical models simulate DQC1 efficiently \cite{animesh}}.  In this model, we start with a known state of an ancilla or probe qubit and couple it to the system. If the system state is known, we can perform spectroscopy of the controlled operation acting  on the system. Else if the operation is known, one can do tomography with the same circuit \cite{scattering}.  In both cases, a measurement performed on the ancilla qubit after the interaction reveals information about the system or the operation. The circuit is given in the figure below
{The circuit diagram for DQC1 is shown below.}
	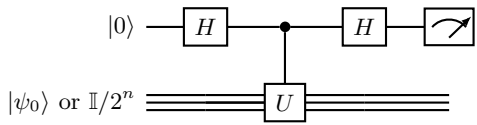
\begin{figure}[!ht]
		\centering
	\begin{quantikz}
		\lstick{$\ket{0}$} & \gate{H} &  \ctrl{1}   & \gate{H} & \meter{} \\
		\lstick{$\ket{\psi_0}$ or $\mathbb{I}/2^{n}$}  & \qwbundle[alternate]{} & \gate{U}  \qwbundle[alternate]{}& \qwbundle[alternate]{} & \qwbundle[alternate]{} 
	\end{quantikz}
\caption{Quantum circuit for the DQC1 protocol (when the input is $\mathbb{I}/2^{n}$). The circuit gives an efficient algorithm for trace estimation of a unitary with only one qubit of quantum information.}
\end{figure}

{The top qubit (the pure qubit that is also the control qubit) is acted upon by a Hadamard gate. This transforms state $\ket{0}$ to $\frac{(\ket{0} +\ket{1})}{\sqrt{2}}$. Then a controlled unitary $U$ is applied followed by another Hadamard gate. It is to be noted that the controlled unitary $U$, and the state 
 $\ket{\psi_0}$ can belong to an arbitrarily large Hilbert space.
Measuring the control qubit, we observe  $\ket{0}$ and  $\ket{1}$ with probabilities}
\begin{align}
P(0)= \frac{1}{2}(1+ \mathtt{Re} \bra\psi_0 U \ket \psi_0)  \nonumber \\
P(1)= \frac{1}{2}(1 - \mathtt{Re} \bra\psi_0 U \ket \psi_0) .
\end{align}

 {Instead of a pure state  $\ket{0}$, if the lower set of qubits are in a completely mixed state, with density matrix, $\rho =  \mathbb{I}/2^{n}$, we get }
\begin{align}
P(0)= \frac{1}{2}(1+ \frac{1}{2^n}\mathtt{Re}  (\mathtt{tr} U))  \nonumber \\
P(0)= \frac{1}{2}(1 - \frac{1}{2^n}\mathtt{Re} (\mathtt{tr} U))   
\end{align}

{By a trivial modification of this scheme, one can make these probabilities depend on $\mathtt{Im} (\mathtt{tr} U)$
and therefore, this gives a quantum algorithm to estimate the trace of a unitary matrix. $L$ measurement of the top qubit will give us an estimate of the trace with fluctuations of size $1/\sqrt{L}$. Therefore, to achieve an accuracy $\epsilon$ one requires $L \sim 1/\epsilon^2$ implementations of the circuit. If $P_e$ is the probability that the estimate departs from the actual value by an amount $\epsilon$, then one needs to run the experiment $L \sim \ln(1/P_e)/\epsilon^2$ times.
This accuracy in the estimate does not scale with the size of the unitary matrix and hence provides an exponential {speed-up} over the best known classical algorithm, provided the unitary admits an efficient gate decomposition. It is known that if the gate decomposition scales as \textit{poly(n)}, the controlled version of these gates also scales \textit{polynomially} in $n$.
Moreover, the result is obtained by a mesaurement of only the top qubit and hence independent of the size of the readout register.}
As a last remark, it is worthwhile to note that, while we have assumed the probe qubit to be in a pure state, this is not necessary. With the probe qubit in a state, $\alpha \ket{0}\bra{0} + \frac{(1- \alpha)}{2}\mathbb{I} $, the model with a tiniest fraction of a qubit is computationally equivalent to the DQC1 circuit described above.
More specifically, the number of runs of the trace estimation algorithm goes as  $L \sim \ln(1/P_e)/\alpha^2\epsilon^2$. Therefore, as long as $\alpha$ is non-zero, the circuit provides an efficient estimate of the trace.


\section{Using DQC1 to calculate OTOC}

\begin{figure*}[!ht]
\centering
 \begin{quantikz}
\lstick{$\ket{0}$} & \gate{H}\slice{$t_1$} &  \ctrl{1} &\qw& \ctrl{1} &\qw& \ctrl{1} &\qw& \ctrl{1} & \gate{H} \slice{$t_2$} & \meter{} \\
\lstick{$\ket{\psi_0}$}&\qwbundle[alternate]{} &\gate{V}\qwbundle[alternate]{}&\gate{U_{\tau}}\qwbundle[alternate]{}&\gate{W}\qwbundle[alternate]{}&\gate{U_{\tau}^\dagger}\qwbundle[alternate]{}&\gate{V^\dagger}\qwbundle[alternate]{}&\gate{U_{\tau}^\dagger}\qwbundle[alternate]{}&\gate{W}\qwbundle[alternate]{}&\gate{U_{\tau}}\qwbundle[alternate]{}&\qwbundle[alternate]{}
\end{quantikz}
\caption{This circuit evaluates the expectation value of OTOC with respect to $\ket{\psi_0}.$ Time progresses along the horizontal line.  The top register is the single-qubit ancilla or probe. The bottom register is the system on which controlled gates act. When the probe qubit is $\ket{0}$, the system is left unchanged, whereas when the probe is $\ket{1}$, controlled operations take place. Measurement of $\sigma_z$ or $\sigma_y$ is performed on the probe qubit, in the end, revealing the value of OTOC. }
\label{fig2}
\end{figure*}
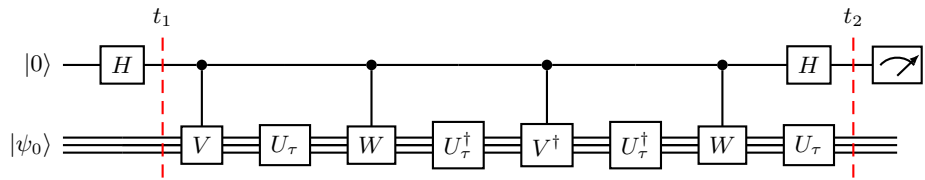

We now adapt the DQC1 algorithm to measure OTOCs. This is shown in the circuit in Fig \ref{fig2}.

Here we initialize the probe to $\ket{0}$ and for simplicity let us say the system state is prepared in a pure state  $\ket{\psi_0}$. The controlled gates act on the system only when the control qubit is $\ket{1}$. $H$ is the Hadamard gate, and $U_\tau$ is the unitary determined by a Hamiltonian which evolves the system up to time $\tau$.  The state of the probe $+$ system at time $t_1$ is $\frac{(\ket{0} +\ket{1})}{\sqrt{2}} \otimes \ket{\psi_0}.$ After the interaction, at time $t_2$, the combined state is $\frac{1}{2}\ket{0} \otimes (1+\mathcal{U}) \ket{\psi_0}+\frac{1}{2}\ket{1} \otimes (1-\mathcal{U}) \ket{\psi_0} $  where $\mathcal{U}= W_\tau^\dagger V^\dagger W_\tau V.$  After the action of the second Hadamard on the probe qubit, measurement of $\sigma_z \otimes \mathbb{I}$, with $\sigma_z$ on the probe qubit yields $\mathtt{Re} \bra{\psi_0}W_\tau^\dagger V^\dagger W_\tau V \ket{\psi_0}$ and measurement of $\sigma_y$ on the probe yields $\mathtt{Im}\bra{\psi_0}W_\tau^\dagger V^\dagger W_\tau V \ket{\psi_0}$. If we perform the circuit sufficiently many times, then we get
\begin{align}
    \langle \sigma_z \rangle &= \mathtt{Re} \bra{\psi_0}W_\tau^\dagger V^\dagger W_\tau V \ket{\psi_0} \nonumber \\ \langle \sigma_y \rangle &= \mathtt{Im} \bra{\psi_0}W_\tau^\dagger V^\dagger W_\tau V \ket{\psi_0}
\end{align}
Thus we have obtained the OTOC values. As mentioned previously, assuming we have an efficient gate decomposition and fix the size of fluctuations in our answer, the complexity if this algorithm does not scale with the dimension of Hilbert space of the physical system under consideration. This is not an unreasonable assumption as efficient decomposition of some quantized chaotic systems is known \cite{benenti2001efficient, emerson2003pseudo, PhysRevA.57.1634} and used in quantum simulations \cite{PhysRevA.68.022302, emerson2002fidelity}.
In the above, the inherent assumption is that the initial state of the system  is perfectly known. By taking the initial state $\ket{\psi_0}\bra{\psi_0}$ to be completely mixed, that is proportional to $\mathbb{I}$, we get the trace of OTOC, which is the measurement with respect to a thermal state at infinite temperature. Therefore, OTOCs with respect to the thermal state at infinite temperature is a perfect candidate for the implementation with DQC1, that employs only 1 qubit of quantum information, and hence a happy accident.

\section{Estimating the eigenvlaue spectrum of OTOC}
\begin{figure*}[!ht]
\begin{quantikz}
\lstick{$\ketbra{0}{0}$} & \gate{H} &  \ctrl{1} &\qw& \ctrl{1} &\qw& \ctrl{1} & \gate{H} &\qw& \meter{} \\
\lstick{$\ketbra{u}{u}$} & \qwbundle[alternate]{}& \gate{FT}\qwbundle[alternate]{} & \qwbundle[alternate]{} &\ctrl{1}\qwbundle[alternate]{}&\qwbundle[alternate]{}& \gate{FT}\qwbundle[alternate]{}& \qwbundle[alternate]{}&\qwbundle[alternate]{}&\qwbundle[alternate]{} \\
\lstick{$\rho_0$} & \qwbundle[alternate]{} &\qwbundle[alternate]{}&\qwbundle[alternate]{}& \gate{W_\tau^\dagger V^\dagger W_\tau V}\qwbundle[alternate]{} &\qwbundle[alternate]{} &\qwbundle[alternate]{}&\qwbundle[alternate]{}&\qwbundle[alternate]{}&\qwbundle[alternate]{}
\end{quantikz}
\caption{Circuit for obtaining the spectral density of OTOC. Now there are two ancillas. Controlled Fourier transform is applied twice on the second ancilla. The operation $W_\tau^\dagger V^\dagger W_\tau V$ which acts on the system is written in a condensed form and should be implemented by decomposing into constituent gates as in Fig. \ref{fig2}. Only the single-qubit probe/ancilla is measured in the end as before.  }
\label{fig3}
\end{figure*}
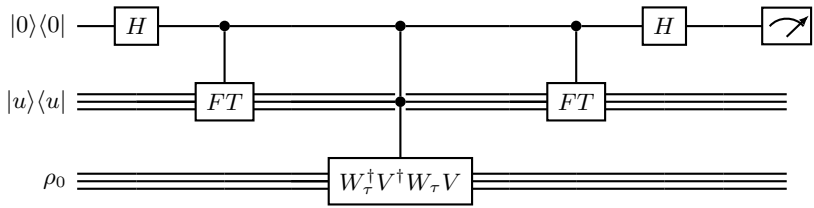

Not only the expectation value of OTOCs, the eigenvalue spectrum of OTOCs is also of interest. Just like  energy eigenvalue spacing for integrable and chaotic systems form distinct distribution,  the level spacing of OTOCs also shows marked difference \cite{spectrum2, spectrum1}. One can obtain the eigenvalue density of OTOCs using a DQC1 algorithm. The circuit is similar to the previous one. But now, apart from the $n$-qubit register for the system,  we also need an extra $n_2$-qubit ancilla  and  perform discrete  Fourier transforms.
The circuit is shown in Fig. \ref{fig3}.

In this circuit, $\ket{u}$ is the initialized state of the second ancilla register of $n_2$ qubits, with the expectation value of OTOC equal to $u$.  The OTOC, $W_\tau^\dagger V^\dagger W_\tau V$, which can be implemented as before.
Let $N_2=2^{n_2}$ and 
at the end of circuit, measuring $\sigma_z$ and $\sigma_y$ on the probe qubit as before, we get
\begin{equation}
f(u)= \frac{1}{N_2} \sum_{s=0}^{N_2-1} \mathrm{exp}(i4 \pi u s/N_2) \mathtt{tr}[(W_\tau^\dagger V^\dagger W_\tau V)\rho_0]
\end{equation}
Where $s$ is the Fourier domain variable of $u.$ Spectral information is now contained in the phases, and can be estimated. Normalizing, so that $\sum_{u=0}^{N_2-1} f(u)=1$,  we get the probability function of eigenvalues. The resolution of the spectral density is determined by the number of ancilla qubits $n_2$. As in the previous case, the DQC1 implementation provides an exponential speed up in obtaining spectral density over any known classical algorithm.

\section{Conclusion}

We have shown that using a single bit of quantum information, one can estimate OTOCs with an exponential speed-up over the best known classical algorithm. In the spirit of the slogan, ``classical chaos generates classical information, as captured by classical Lyapunov exponents and the classical Kolmogorov-Sinai entropy, quantum chaos generates quantum information", leading to the growth of OTOCs (till the Ehrenfest time), {which are} popular quantifiers for this. In this work, we have given an efficient quantum algorithm for estimating OTOCs and capturing the growth of quantum complexity.
One possible avenue is to estimate the semiclassical 
formulas, like the Gutzwiller trace formula on a quantum computer. 
There are existing algorithms for this \cite{georgeot08} that give a 
polynomial speed-up over similar implementations on a classical 
computer. 
We aim to explore the possibility of such computations using the DQC1 model of quantum computation, which can even operate on highly mixed initial states. One can also consider a perturbed OTOC where the operator $W_\tau^\dagger$ that occurs in  $W_\tau^\dagger V^\dagger W_\tau V$,
 undergoes time evolution with a slightly perturbed Hamiltonian as compared to  $W_\tau$ and therefore provides a direct analog to classically chaotic systems under stochastic noise. {
Moreover, understanding the power behind DQC1 is still an open question.
 Future directions include determining the nature of resources 
quantum mechanics provides for information processing tasks that are 
superior to their classical counterparts as well as other avenues where mixed-state quantum computation can be applied.}
\bibliography{reference_otoc}

\end{document}